# Ferroelectric Exchange Bias Affects Interfacial Electronic States

*Gal Tuvia [1], Yiftach Frenkel [2], Prasanna K. Rout [1], Itai Silber [1], Beena Kalisky [2†] and Yoram Dagan [1‡]*

1. Raymond and Beverly Sackler School of Physics, Tel Aviv University, Tel Aviv 6997801, Israel

2. Department of Physics and Institute of Nanotechnology and Advanced Materials, Bar-Ilan University, Ramat-Gan 5290002, Israel

† E-mail: beena@biu.ac.il

‡ E-mail: yodagan@tauex.tau.ac.il

**Abstract:**

In polar oxide interfaces phenomena such as conductivity [1], superconductivity [2,3], magnetism [4–9], one-dimensional conductivity [10,11] and Quantum Hall states [12] can emerge at the polar discontinuity. Combining controllable ferroelectricity at such interfaces can affect the superconducting properties and shed light on the mutual effects between the polar oxide and the ferroelectric oxide. Here we study the interface between the polar oxide $LaAlO_3$ and the ferroelectric Ca-doped $SrTiO_3$ by means of electrical transport combined with local imaging of the current flow with the use of scanning Superconducting Quantum Interference Device (SQUID). Anomalous behavior of the interface resistivity is observed at low temperatures. The scanning SQUID maps of the current flow suggest that this behavior originates from an intrinsic bias induced by the polar $LaAlO_3$ layer. Our data imply that the intrinsic bias combined with ferroelectricity constrain the possible structural domain tiling near the interface. We recommend the use of this intrinsic bias as a method of controlling and tuning the initial state of ferroelectric materials by design of the polar structure. The hysteretic dependence of the normal and the superconducting state properties on gate voltage can be utilized in multifaceted controllable memory devices.

The interfaces between polar and non-polar oxides exhibit unique electronic properties that are conveniently controllable by applied gate voltage. Understanding the electronic structure of these interfaces is crucial for mastering them and for implementing them in oxide-based electronics [13]. For example, the hallmark interface between $LaAlO_3$ and $SrTiO_3$ (LAO/STO) exhibits a threshold for the conductivity of four LAO epitaxial unit cells [1]. This critical behavior has been explained as the result of electronic reconstruction due to the polar field of the LAO layer, which is also accompanied by lattice distortion on the STO side [14], as observed by x-ray diffraction [15,16] and scanning transmission microscopy experiments [17,18].

STO is a band insulator with a perovskite cubic structure at room temperature. At ~105 K the oxygen octahedron antiferrodistortively rotates around one of its principle axis [19]. Pristine STO is a quantum paraelectric [20]. However, upon being doped with small amounts of Ca or by substituting $O^{18}$ for $O^{16}$, the ferroelectric transition recovers with a ferroelectric Curie temperature that depends on the concentration of Ca [21] or $O^{18}$ [22,23]. Creating oxygen vacancies, or substituting Sr by La or Ti by Nb, can turn STO into a conductor and even a superconductor with transition temperature nonmonotonically depending on doping. It has been shown that superconductivity can exist in the ferroelectric-like bulk of doped STO and can even be enhanced by introducing ferroelectricity [24–30].

Here we explore a new route for realizing a 2D polar metal that can become a superconductor at the interface between a 3D ferroelectric insulator ($Sr_{1-x}Ca_xTiO_3$ with x=0.01, 0.0025, 0.002) and a polar oxide ($LaAlO_3$). We find tunable 2D superconductivity along with anomalous behavior of the resistivity at low temperatures. To understand this behavior, we combine transport and scanning SQUID (Superconducting Quantum Interference Device) measurements to map the spatial distribution of the current flow as a function of gate voltage below the ferroelectric transition temperature. Our data suggest that the top polar layer exerts an effective negative bias (carrier depletion) on the ferroelectric material near the interface. This effective bias is somewhat analogous to the exchange bias often used in magnetic devices [31]. The ferroelectric polarization in the bulk is switchable by an external gate voltage, resulting in a hysteretic sheet resistance and superconducting critical temperature. This memory effect has a controllable initial state by design of the polar structure.

An anomalous sharp increase in the $LaAlO_3/Sr_{0.99}Ca_{0.01}TiO_3$ interface sheet resistance $R_\square$ is observed below the ferroelectric transition temperature, ~30K, followed by superconductivity below 300 mK (see **Figure 1**a). Ferroelectricity is also demonstrated in **Figure 1**b where we show the temperature evolution of reproducible hysteresis loops in the resistance versus gate-voltage characteristics. These loops close as we increase the temperature towards the ferroelectric transition. Ferroelectricity also affects the superconducting properties of the interface, as we show in the inset of **Figure 1**a. The superconducting transition temperature *at zero gate voltage* shifts after being approached from either +20V or –20V, demonstrating control of the 2D interface superconductivity by the hysteretic behavior of the 3D ferroelectric bulk.

To understand the anomalous resistance increase at low temperatures and the microscopic details of the current flow, we performed scanning SQUID mapping of the device sketched in **Figure 1**c. The current produces magnetic flux, which is captured by the SQUID pick-up loop, enabling us to evaluate the details of the 2D current distribution at the interface [32] (Experimental Section). Scanning SQUID images of the sample zero-electric-field-cooled to 4.2K show that the current occupies only part of its entire lithography-available channel (**Figure 1**d left). The current flow occupies the entire available width only when a positive gate is applied. (**Figure 1**d right).

**Figure 2**a shows the detailed gate-voltage dependence of the current-flow pattern at 4.2K in our sample as revealed by scanning SQUID measurements. For samples cooled at zero electric field (no gate voltage applied) the current flow occupies only 80% of the 100 microns available channel as defined by lithography (inset of **Figure 2**a). Upon applying positive gate voltage (accumulating electrons) the width of the current path increases, reaching saturation at ~15V when it fully occupies the entire lithography-defined 100-micron width. Similarly, applying negative gate voltage (depleting electrons) narrows the current path further (Supporting Information **Figure S3**). A similar effect of negative-gate application has also been demonstrated for the LAO/STO interface [33]. However, the initial zero-gate narrow path is unique to the ferroelectric $LaAlO_3/Ca_{0.01}Sr_{0.99}TiO_3$ (LAO/CSTO) interface.

We conclude that when the sample is cooled to below the ferroelectric transition temperature, an effective negative gate bias (electron depletion) is built. With the use of the 15V required to fully occupy the channel and the capacitance (**Figure S2**), one can estimate the total (all electronic bands) carrier depletion to be $3 \times 10^{12}$ cm$^{-2}$. This estimate is a lower boundary for the real number as the device's effective capacitance per unit area can be greater than the parallel-plate approximation [34]. We note that, on the assumption of constant carrier density, the decrease of 20% of the current path can account for merely a 25% increase in resistance seen in **Figure 1**a as the sample is cooled below 30K to 0.3K.

A broad-view scanning SQUID image of our zero-electric-field-cooled device is shown in **Figure 2**b. The current density is somewhat greater at the twin boundaries appearing as features tilted by 45º relative to the crystal axis (see Supporting Information **Figure S**10 for more information). It has been shown, that for STO-based heterostructures, the current flow is modulated along the structural twin boundaries [35]. We observe a clear tendency for the formation of diagonal twin boundaries at 45º with respect to the crystal axis, as shown in this scan (Supporting Information, **Table S1**). We suggest that the ferroelectric polarization near the interface favors pointing perpendicular to the interface (along the Z axis), parallel to the LAO internal dipole moments (see illustration in **Figure 4**a). Since the polarization of the ferroelectric CSTO is perpendicular to the crystallographic c-axis (axis of rotation of oxygen octahedra) [21], X and Y structural domains should be abundant at the interface. Boundaries between X and Y domains result in diagonal (45°) twin boundaries [36], consistent with our experimental observation.

We also note that Honig *et al.* [36] reported a reduction in the number of Z domains for the standard LAO/STO interface when a large negative gate voltage is applied, consistent with our interpretation of the ferroelectric exchange-bias effect.

The hysteresis in resistance versus gate voltage can become very large depending on gate-voltage history and temperature. We show such an example in **Figure 3**a when strong, reproducible hysteretic behavior is observed in the resistance measured at 600mK after gate training (Same sample as in Figure 1b, see Supporting Information for more details). We note that this behavior is very different from that of the non-ferroelectric (standard) LAO/STO interfaces where only the initial gate scan is different from the successive scans, which are then independent of the gate-sweep direction as long as the maximum voltage has not been exceeded [37].

At $T$~300mK, a superconducting transition is observed. Ferroelectric hysteresis still plays a role at the superconducting regime as displayed in **Figure 3**b, where superconducting $T_c$ changes for different gate-sweep directions. Our results show that an extremely large response can be achieved at a fixed temperature and zero applied gate by merely sweeping the gate voltage up or down and then back to zero (resistance changing from zero for the higher $T_c$ state to a finite value for the lower $T_c$ state). We note that the superconducting $T_c$ depends only on carrier density [38] while the sheet resistance depends also on the current-flow details as shown by scanning SQUID measurements (see Supplementary **Figure S3** for more details on how the gate voltage affects the current pattern) and on the mobility, which itself depends strongly on gate voltage [39]. We also note that superconducting $T_c$ seems to be saturated for the positive gate voltage regime (up to 300V, see Supplementary **Figure S4**) in contrast to the "dome"-shaped superconducting region at the standard LAO/STO interface [39].

To further study the superconducting properties, we show in **Figure 3**c and **Figure 3**d respectively the superconducting out-of-plane and in-plane critical magnetic fields with respect to temperature. The out-of-plane critical field exhibits a linear temperature dependence while the in-plane critical field follows a square-root temperature dependence as expected for a two-dimensional superconductor. The in-plane critical field exceeds the Clogston-Chandrasekhar limit indicative of the strong spin-orbit interaction in the Ti bands [40].

Our understanding of the polar and structural configurations of the system is depicted in **Figure 4**. **Figure 4**a describes the polar structure below the ferroelectric transition temperature at zero gate voltage. The direction of polarization near the interface is constrained to be parallel to the LAO internal dipole moments. This constrained polarization near the interface results in the measured effective negative bias. The effect of the interface decays further into the bulk, where average net polarization is zero (see example for domain configuration in typical bulk ferroelectric perovskite [41]). Applying an external electric field (gate voltage) controls polarization in the bulk in a hysteretic fashion, characteristic of ferroelectric materials. **Figure 4**b and **Figure 4**c explain how the diagonal (45°) twin boundaries (see Figure 2b) are a consequence of the Z-pointing polarization. How can we explain the strong increase in resistance upon cooling from the ferroelectric transition to the superconducting one? We relate this increase to four correlated effects resulting from the intrinsic negative bias: a. narrowing of the current path, b. carrier depletion, c. mobility reduction, d. ferroelectric constraints on domain tiling.

The first two simply increase the resistance. It has been shown for the standard LAO/STO interface that carriers fill the Ti bands a few unit cells away from the interface. A negative bias (or in our case, the exchange bias coming from the ferroelectric polarization) pushes the mobile electrons towards the interface thus reducing their mobility [39,42]. This effect may also apply to our devices and further increase the resistance as the effective bias is built-up upon cooling. It has also been shown that twin boundaries are highly conductive at the LAO/STO interface [35]. Our data suggest that the intrinsic bias results in preferred diagonal X/Y twin boundaries. This diverts some of the current flow for lithography-defined channels patterned along the x or y directions (see Supplementary **Figure S10**). This view is supported by the fact that for large devices measured in a Van der Pauw configuration a much smaller effect is observed (see Supplementary **Figure S5** a).

We note that when 3D bulk conductivity is induced in Ca-doped STO by the creation of oxygen vacancies (obviously without any intrinsic bias), only a small change in the resistance is observed below the ferroelectric transition temperature [24,43], further demonstrating the importance of the intrinsic bias and current confinement for observing such a large effect. Ferroelectric control of the LAO/STO interface was also demonstrated by polarizing a ferroelectric layer deposited on the LAO [44]. We also note that a polar 2D metal has been realized in the $BaTiO_3/SrTiO_3/LaTiO_3$ heterostructure [45]. Our system is therefore unique since a tunable 2D polar metal that can become a superconductor is created at the interface between a 3D ferroelectric bulk and a polar insulator. Ferroelectric materials tend to be insulators and were previously believed to exist in a separate domain from superconductivity [46]. Nevertheless, superconductivity and ferroelectricity can coexist in some rare cases [47], and it has been recently suggested that ferroelectric fluctuations may even increase superconducting $T_c$ [24–29]. We do not report any significant increase of superconducting $T_c$, even for calcium concentrations of 0.2% and 0.25%, which are close to the ferroelectric quantum critical point where fluctuations are strongly enhanced (Supporting Information, **Figure S7**). It is possible that the effective bias induced by the LAO constrains the direction of polarization near the interface so that quantum fluctuations are presumably less effective and consequently the assumed $T_c$ enhancement is not observed.

Finally, we note that the critical thickness for conductivity for the LaAlO$_3$/Sr$_{1-x}$Ca$_x$TiO$_3$ (x=0.01,0.0025,0.002) heterostructure is reduced to three unit cells of LAO in contrast to the four-unit-cell threshold reported for the standard LAO/STO interface (see Supplementary **Figure S6**). We conjecture that the small calcium ion makes the lattice distortions associated with the charge transfer to the interface [15–18] easier and consequently reduces the critical thickness.

To summarize, we conclude that polar-oxide layers exert an internal electrical bias, orienting the ferroelectric polarization near the interface parallel to the polar-oxide internal dipole moment, thus strongly affecting the electronic properties of conducting oxide interfaces. This effect is analogous to the magnetic exchange bias. We come to this conclusion with the use of the following experimental evidence: (a) An anomalous increase in interface resistance as the temperature is decreased to below the ferroelectric transition temperature, (b) Using scanning SQUID current imaging, we find that the current does not occupy its entire available channel. Upon applying positive bias, the current fully occupies the lithography-defined current path, (c) Abundance of X-Y twin boundaries indicative of a preferred polarization along the Z direction. While the polarization near the interface is constrained, polarization further away into the bulk is switchable by externally applying an electric field. This translates to a hysteretic behavior of the interfacial resistance and its superconducting $T_c$ with respect to the applied gate voltage. The strong hysteresis of the resistance above the superconducting-transition temperature and the superconducting-memory effect can be utilized in future superconducting-memory devices.

**Experimental Section**

*Sample preparation and transport measurements*:

Epitaxial layers of LaAlO$_3$ were grown with the use of the Pulsed Laser Deposition (PLD) technique monitored by reflection high-energy electron diffraction (RHEED) at a partial pressure of oxygen of $1\times10^{-4}$ Torr, and a temperature of 780°C, as described in reference [5]. The layers were deposited on atomically flat TiO$_2$ terminated, 0.5 mm thick {100} Sr$_{1-x}$Ca$_x$TiO$_3$ (with x=0.01, 0.0025 and 0.002) substrates. The resulting interface is conducting for LAO thicknesses of three or more unit cells, in contrast to the four-unit-cell threshold of the non-ferroelectric (standard) LAO/STO interface. Au back-gate electrodes were attached to the bottom of the CSTO. The capacitance between the Au and the conducting interface has ferroelectric characteristics (see Supplementary **Figure S2**). When gate voltage was applied, the leakage current was immeasurably small (<1pA). The gate voltage is defined as positive when electrons accumulate at the interface. Measurements were performed in a dilution refrigerator with a base temperature of 20mK under magnetic fields of up to 8T. SQUID and superconductivity data were measured with the use of a 100µm×700µm Hall bar device suitable for both SQUID and transport measurements. The current path was defined with the use of an amorphous material hard mask (see ref. [48] for details). Both van der Pauw and Hall configurations showed similar results, differing only in the magnitude of the resistance increase below the ferroelectric transition temperature and response to gate voltage because of their different geometries [34]. Transport results were reproduced for 3 and 10 unit cells of LaAlO$_3$ thick samples. All measurements shown in this paper are for the LaAlO$_3$/Sr$_{0.99}$Ca$_{0.01}$TiO$_3$ interface with 10 unit cells of LAO. Data for different dopings of Ca and for 3 unit cells of LAO are presented in the Supporting Information. SQUID scans were conducted for several samples with 1%Ca concentration. Five of the six devices showed only diagonal domains while one device showed 90 and 0 domains. The initially narrower current path and its widening by the use of positive gate voltage was reproducibly measured. Hysteresis measurements of resistance vs gate voltage, presented in **Figure 1**b and **Figure 3**a, were conducted with a large 5x5mm$^2$ Van der Pauw sample to demonstrate gate control. All other SQUID and transport measurements presented in this paper were conducted with the same 100-micron-wide device.

*Scanning SQUID measurements*:

The local measurements were carried out with the use of a custom-built piezoelectric-based scanning SQUID microscope with a 1.8mm diameter pick-up loop [49,50]. The scanning SQUID microscope was used to image magnetic flux generated by current flow at the samples as a function of position. The measured flux is given by $\phi_s = \int g(x,y) B \cdot \vec{da}$ where the integral is taken over the plane of the SQUID, $g(x,y)$ is the point-spread function of the pickup loop, $B$ is the magnetic field that originated from the sample and $\vec{da}$ is the infinitesimal area vector element pointing normal to the plane of the SQUID. The measurements were performed by applying an AC current to the sample and collecting the flux created by currents in the sample using lock-in techniques allowing ~$10^{-6}$ $\Phi_0$ flux sensitivity where $\Phi_0$ is the magnetic flux quantum. Each flux image is a convolution of the z component of the magnetic field and the SQUID point-spread function. A current-carrying wire will appear in our images as red stripes next to blue stripes indicating the positive and negative magnetic-field lines circling the wire.

**Acknowledgements**

G.T. and Y.F. contributed equally to this work. We acknowledge Ekhard Salje, Margherita Boselli, Gernot Scheerer, Jonathan Ruhman and Lior Kornblum for useful discussions.

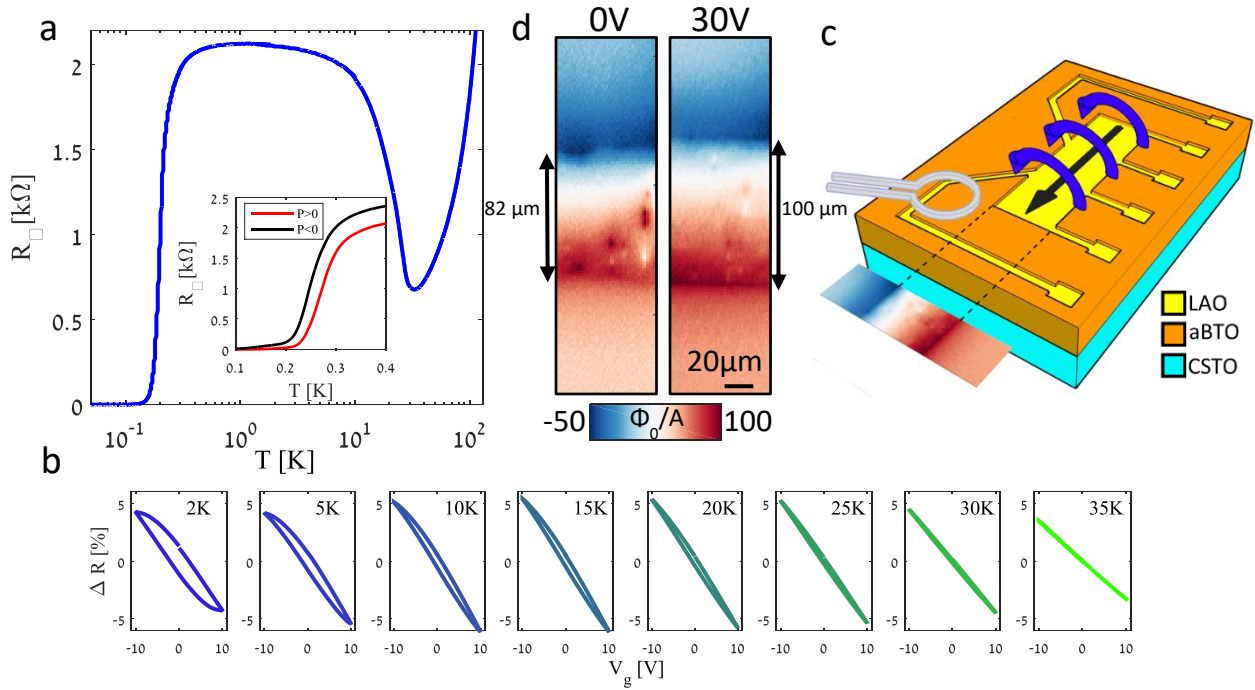

**Figure 1:** a) Sheet resistance of the LaAlO$_3$/Sr$_{0.99}$Ca$_{0.01}$TiO$_3$ interface versus temperature on a logarithmic scale. A clear increase of resistance is visible below the ferroelectric transition temperature ~30K, followed by superconductivity below 300mK. Inset: Superconducting transitions for zero electric field (no gate voltage) for different gate-sweep directions display different superconducting critical temperatures, demonstrating control of the 2D interface superconductivity by the hysteretic behavior of the 3D ferroelectric bulk. b) Sheet resistance responds to gate hysteretically, the hysteresis loop increases as the temperature is lowered below the ferroelectric critical temperature, above which no hysteresis is observed. c) Illustration of the system: Amorphous BaTiO$_3$ (aBTO orange) defines the insulating regions confining the current. The LaAlO$_3$/Sr$_{0.99}$Ca$_{0.01}$TiO$_3$ interface regions (LAO/CSTO light blue/yellow) are conducting and can be probed by the SQUID. Black arrow symbolizes current flow through the channel. Blue arrows indicate magnetic field lines probed by the SQUID pick-up loop (grey). A typical SQUID scan is shown with the lithography-defined current width marked by dashed lines. The color code for magnetic flux is also presented. d) SQUID scans of the LaAlO$_3$/Sr$_{0.99}$Ca$_{0.01}$TiO$_3$ interface at 4.2K. Current occupies only about 80% of the entire lithographically-defined channel width for sample cooled at zero electric field (left). When positive gate is applied, the current flows in the entire lithographically-defined channel width (right). Black arrows mark the average width of the current path. These results indicate that the channel is cooled with an intrinsic effective negative bias.

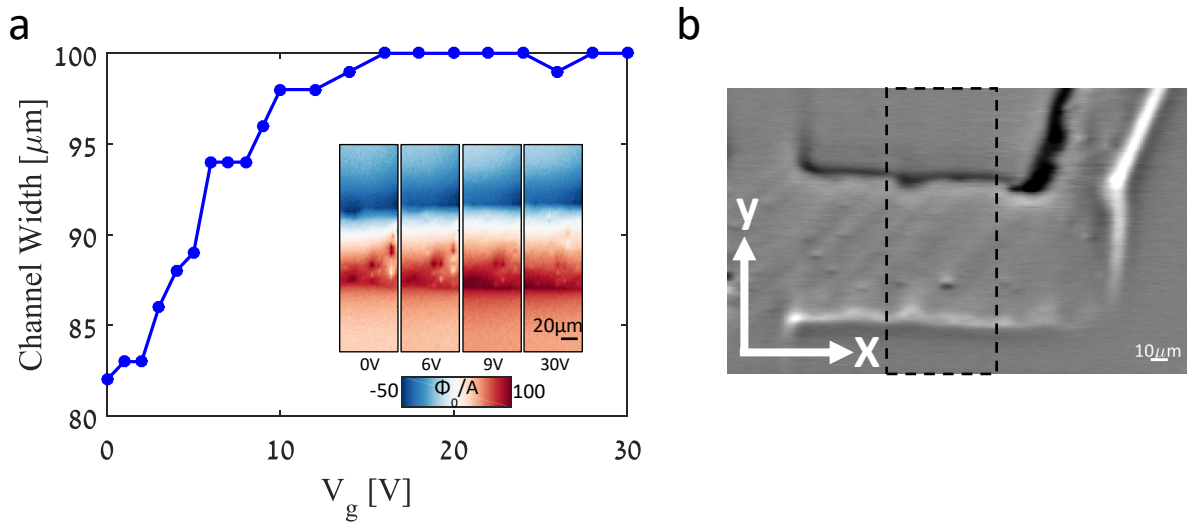

**Figure 2:** a) Current width through the 100-micron lithographically-defined channel measured by scanning SQUID at 4.2K. The width of the current flow pattern increases as positive gate voltage is applied. For samples cooled at zero electric field, the current occupies only 80% of the lithographically-defined channel path. We relate this to an intrinsic effective bias which is a result of the interaction between the polar LaAlO$_3$ and the ferroelectric Sr$_{0.99}$Ca$_{0.01}$TiO$_3$ (1% CSTO). This effective bias is compensated by the application of a positive gate voltage of ~15V. Inset: SQUID scans at selected gate voltages used to extract the width of the current-flow pattern. In addition to the narrowing current path, the zero-field cooled sample exhibits spots with reduced current flow. These spots gradually disappear when positive gate is applied. b) A wider SQUID scan of our device shows current modulations along domain walls as can be seen by the 45-degree stripe pattern. Dashed rectangle marks the scanning area of **Figure 2**a and **Figure 1**d. White arrows denote the X and Y crystal axis.

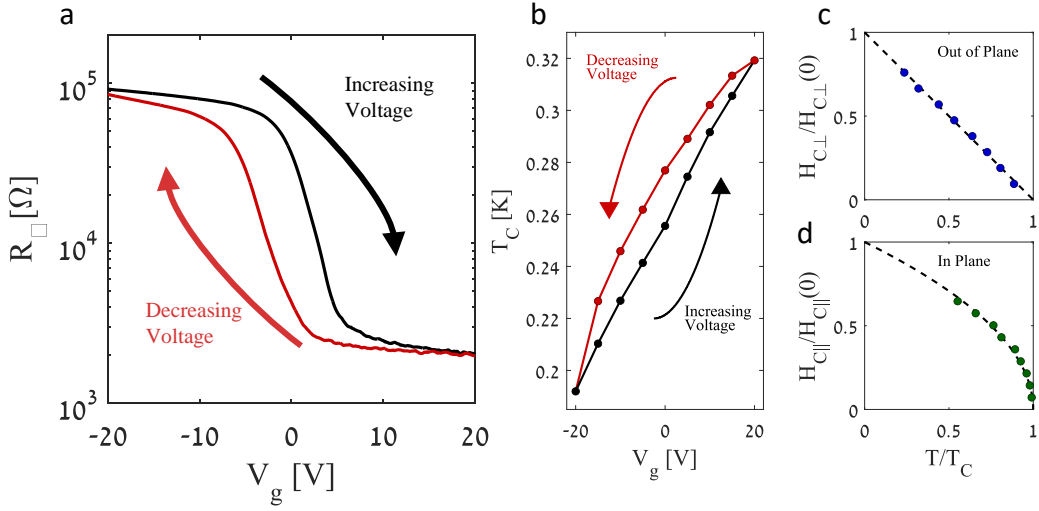

**Figure 3:** a) Hysteresis loop of sheet resistance versus gate voltage at 600mK. b) Hysteresis in superconducting $T_c$. c) and d) Out-of-plane and in-plane critical fields respectively follow linear and square-root dependences on temperature, typical of 2D superconductors. $H_{c\parallel}(0)$ and $H_{c\perp}(0)$ are the critical fields extrapolations for zero temperature. We obtain $H_{c\parallel}(0)=3.5[T]$, $H_{c\perp}(0)=0.26[T]$. These measurements were taken at gate voltage of 20V. $H_c$ (and $T_c$) is defined as the field (temperature) at which the resistance is half of that at 650mK.
10

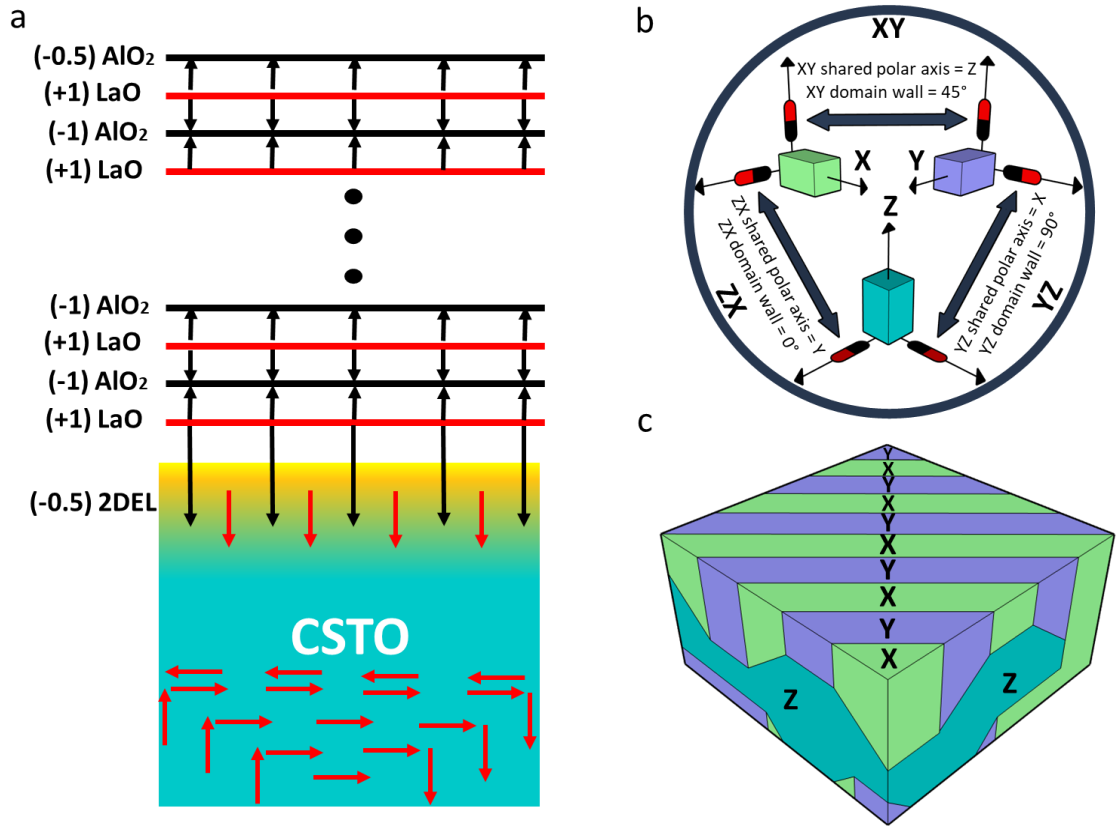

**Figure 4:** a) Illustration of our understanding of the polarization in the LAO/CSTO interface: Ferroelectric polarization is marked with red lines. Polarization near the interface has a preferential alignment as a result of the effective bias exerted by the LaAlO₃. Polarization further away from the interface is oriented in some unknown way with zero net polarization. Black arrows indicate electric field. b) The ferroelectric polarization in Ca-doped SrTiO₃ is perpendicular to the structural c-axis. Each pair of domains (XY, YZ, ZX) creates a domain wall with a specific angle with respect to the crystal axis (45°, 90°, 0° respectively). For ferroelectric $Sr_{0.99}Ca_{0.01}TiO_3$, two domains sharing a domain wall have also a possible common axis of polarization (XY, YZ and ZX domain pairs may only have similar polarizations along the Z, X and Y axis respectively). c) The polarization of ferroelectric domains near the interface prefer pointing down the Z direction, resulting in preferred diagonal XY domain boundaries at the interface. The domains further away into the bulk are assumed to return to their unperturbed distribution.